\documentclass[preprint,11pt,numbers,sort&compress]{elsarticle}

\usepackage{amsmath,amssymb, amsthm}
\usepackage{siunitx}
\usepackage{todonotes}
\usepackage{float}
\usepackage{enumitem}

\newcommand{\fieldBPF}{E_{\mathrm{BFP}}}

\newcommand{\xb}{\vec{x}_b}
\newcommand{\xf}{\vec{x}_f}
\newcommand{\panel}[1]{\textbf{(#1)}}

\journal{\hspace{-55pt}\fcolorbox{white}{white}{\phantom{XXXXXXX}}}

\begin{document}

\begin{frontmatter}


\title{Interactive Simulation and Visualization of Point Spread Functions in Single Molecule Imaging}

\author[hhmi,tuwien]{Magdalena C. Schneider}            
\author[jku]{Fabian Hinterer}
\author[meduniinnsbruck]{Alexander Jesacher}
\author[tuwien]{Gerhard J. Schütz}

\affiliation[hhmi]{organization={HHMI Janelia Research Campus},
            addressline={19700 Helix Drive}, 
            city={Ashburn},
            postcode={20147}, 
            state={VA},
            country={USA}}

\affiliation[tuwien]{organization={TU Wien, Institute of Applied Physics},
            addressline={Lehargasse 6}, 
            city={Vienna},
            postcode={1060}, 
            country={Austria}}

\affiliation[jku]{organization={JKU Linz},
            addressline={Altenbergerstraße 69}, 
            city={Linz},
            postcode={4040}, 
            country={Austria}}
            
\affiliation[meduniinnsbruck]{organization={Medical University of Innsbruck},
            addressline={Müllerstraße 44}, 
            city={Innsbruck},
            postcode={6020}, 
            country={Austria}}

\begin{abstract}
The point spread function (PSF) is fundamental to any type of microscopy, most importantly so for single-molecule localization techniques, where the exact PSF shape is crucial for precise molecule localization at the nanoscale. However, optical aberrations and fixed fluorophore dipoles can lead to non-isotropic and distorted PSFs, thereby complicating and biasing conventional fitting approaches. In addition, some researchers deliberately modify the PSF by introducing specific phase shifts in order to provide improved sensitivity, e.g.,\ for localizing molecules in 3D, or for determining the dipole orientation. For devising an experimental approach, but also for interpreting obtained data it would be helpful to have a simple visualization tool which calculates the expected PSF for the experiment in mind. 
To address this need, we have developed a comprehensive and accessible computer application that allows for the simulation of realistic PSFs based on the full vectorial PSF model. It incorporates a wide range of microscope and fluorophore parameters, enabling an accurate representation of various imaging conditions. Further, our app directly provides the Cramér-Rao bound for assessing the best achievable localization precision under given conditions.
In addition to facilitating the simulation of PSFs of isotropic emitters, our application provides simulations of fixed dipole orientations as encountered, e.g.,\ in cryogenic single-molecule localization microscopy applications.
Moreover, it supports the incorporation of optical aberrations and phase manipulations for PSF engineering, as well as the simulation of crowded environments with overlapping molecules. 
Importantly, our software allows for the fitting of custom aberrations directly from experimental data, effectively bridging the gap between simulated and experimental scenarios, and enhancing experimental design and result validation.

\end{abstract}

\end{frontmatter}



\section{Introduction}
Single-molecule localization microscopy (SMLM) techniques offer a powerful approach to discern molecular structure and dynamics of biological samples below the diffraction limit \cite{lelek2021}. The overall idea is to virtually dilute single molecule signals in time, e.g.,\ by stochastic switching of fluorophores between a bright and a dark state \cite{Rust_2006, betzig2006}, by transient binding of fluorescent ligands \cite{sharonov2006, jungmann2010}, or by massive underlabeling in single molecule tracking \cite{wieser2008}. All these methods yield image stacks with very low densities of single molecule signals per frame, which ideally show no overlap. From such images it is possible to estimate the emitter positions to a precision that is mainly limited by the signal to noise ratio of the single molecule images \cite{smith2010}.

Crucially, the reliability of the final superresolution images or acquired tracking data critically depends on the quality of the positions obtained from the fitting procedure. In the simplest case, single molecule signals are fitted using a Gaussian function \cite{Huang_2008}. For this, it is assumed that a fluorophore's emission yields an isotropic point spread function (PSF). However, various factors such as microscope aberrations and fluorophore characteristics alter the shape of the PSF. The mismatch of the fitted simple model of a Gaussian function and the intricate shape of the true PSF may well lead to biases in the estimated positions of up to tens of nanometers, thus highly distorting the obtained results \cite{engelhardt2011}. Hence, it is vital to incorporate a realistic model in the fitting procedure \cite{HinSch_2022}.

In response to this challenge, we have developed a comprehensive and accessible computer application that allows for interactive simulation and visualization of realistic PSFs under a wide range of microscope and fluorophore parameters, enabling an accurate representation of various imaging conditions. In contrast to similar previous tools that allow for calculating and displaying PSFs \cite{Kirshner_2013}, we implement the full vectorial PSF model \cite{Mortensen_2010, Aguet_2009, Axelrod_2012} and allow to set a fixed fluorophore dipole orientation; the application is hence not limited to PSFs of isotropic emitters.

In practice, fluorophores are dipole emitters, and their rotation is often restricted by steric hindrances \cite{hulleman2021}. Recently, researchers became interested in performing SMLM under cryogenic temperatures \cite{weisenburger2017, li2015}. Under these conditions, the dipole orientation of a fluorophore is fixed, which leads to generally anisotropic emission patterns and altered PSFs \cite{Enderlein_2006}. In addition to facilitating the simulation of isotropic PSFs, our tool therefore allows to particularly simulate and visualize the PSF of non-rotating anisotropic emitters with arbitrary dipole orientations.

The best precision that can be achieved in localization techniques depends on the signal to noise ratio, the PSF shape, and the chosen fitting procedure. Ultimately, the localization precision is limited by the Cramér-Rao bound (CRB) \cite{Kay_1993}, which is a theoretical limit for the precision any unbiased estimator can possibly achieve under the given conditions. Notably, the CRB depends on the shape of the PSF. This fact can be taken advantage of by shaping the PSF via manipulations in the back focal plane, often referred to as PSF engineering \cite{Khare2023,lelek2021}. First, this allows to shape the PSF in a way to achieve best lateral position estimate. Second, this breaks the ground for estimating not only lateral position, but also encoding additional parameters including axial position and dipole orientation in the PSF \cite{hulleman2021}. To easily assess the effect of different PSF engineering approaches, we included a feature for adding manipulations of the emitted light in the back focal plane in our app. Eventually, we provide the option to directly determine the CRB for any given PSF shape.

In practice, the shape of the PSF is often affected and degraded by various types of aberrations, for example imperfections of the optical setup such as coma, spherical aberrations and astigmatism. These aberrations negatively affect the fitting procedure and lead to decreased quality of the position estimates and all other estimated parameters. In our app, arbitrary aberrations can be easily included using Zernike coefficients. Moreover, we provide an extended feature that allows for fitting a specific microscope’s aberrations from data recorded from a calibration sample. The obtained coefficients can be directly loaded back into the PSF simulation app, yielding simulation results tailored to the user’s setup.

As a final feature, we implemented the option to simulate multiple fluorophores in the same region of interest. Overlapping PSFs often occur in cryoSMLM applications, as switching of fluorophores between bright and dark states is decelerated under these conditions \cite{li2015,hulleman2018}. In addition, overlapping PSFs occur in step-wise photobleaching methods \cite{Gordon_2004}, where the signals of multiple fluorophores initially overlap, complicating the localization procedure.

Our application allows to visualize and examine the single fluorophore images, as they would be obtained for specific conditions. It further allows to export the obtained images for further analysis. It could be used, e.g.,\ to provide ground truth information for more advanced fitting procedures. It supports both the import and export of data, encompassing input parameters and the calculated PSF images. This facilitates a streamlined and accurate approach for the assessment of PSFs under various conditions and PSF engineering approaches.


\section{Features}
\begin{figure}[!htb]
    \centering
    \includegraphics[width=\textwidth]{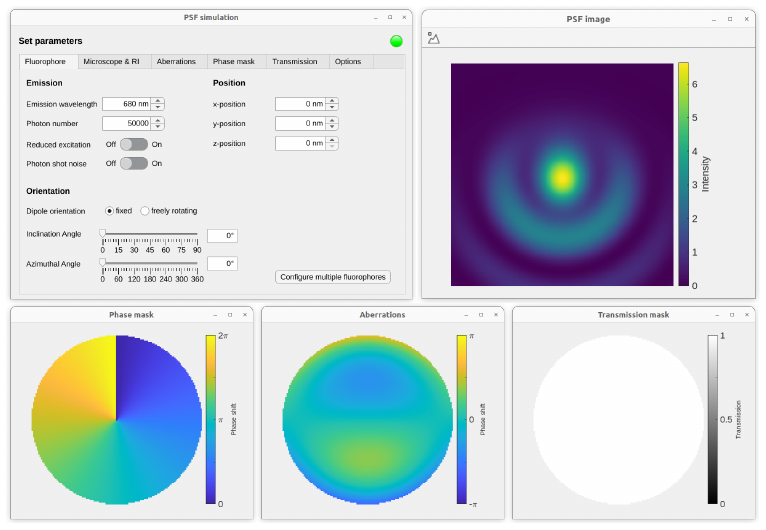}
    \caption{Overview of app windows. The main app window (top left) comprises several tabs and allows to set all simulation parameters and visualization options. The PSF is visualized in a separate window (top right). The bottom row shows the applied phase mask (left), the aberration (middle) and transmission mask (right).}
    \label{fig:windows_overview}
\end{figure}

\begin{figure}[!ht]
    \centering
    \includegraphics[width=0.7\textwidth]{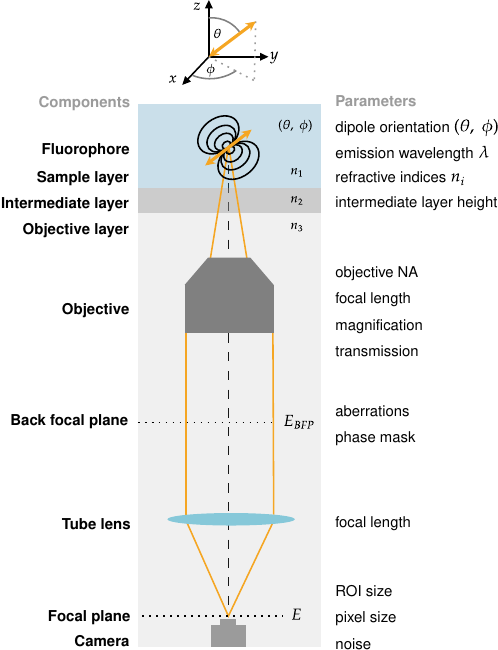}
    \caption{Schematic of implemented microscope setup. \panel{a} On the left, the components of the microscope are listed; on the right, all the parameters that can be varied in the simulations. $E_\text{BFP}$ and $E$ are the electric fields in the back focal plane of the objective and in the focal plane of the tube lens, respectively. The path of the ray is simplified and intended to show the infinity space between the objective and the tube lens where phase manipulations can be easily modeled. \panel{b} Inclination angle $\theta$ and azimuthal angle $\phi$ defining the dipole orientation }
    \label{fig:setup}
\end{figure}

Our application is structured into several subwindows (Fig.~\ref{fig:windows_overview}). The main window allows to set all simulation parameters and visualization options. Fig.~\ref{fig:setup} shows a schematic overview of the implemented setup and all parameters that can be varied. We simulate the emission of a fluorophore as a dipole emitter and its PSF as observed in a conventional wide-field microscope setup \cite{Axelrod_2012}. Here, we will present a short overview of the most important features and illustrative examples of obtained PSFs. For an exhaustive documentation detailing all parameter settings and screenshots of all tabs and subwindows of the app we refer to the app manual in the Supplementary Material.

\label{section:features}

\subsection{Fluorophore and microscope parameters}
First, the app allows to configure the sample and microscope. The sample is assumed to be a fluorophore, modeled as dipole emitter \cite{Axelrod_2012}.
The fluorophore parameters that can be set include the emission wavelength, its lateral and axial position, and the number of observed photons. Further, the fluorophore can be assumed to be either freely rotating, yielding an isotropic PSF, or fixed. In the latter case, the dipole orientation of the fluorophore can be specified via its inclination angle $\theta$ and azimuthal angle $\phi$. In addition, the app allows to account for reduced excitation due to dipole inclination by automatically reducing the number of observed photons. For simplicity, the excitation dipole is assumed to be aligned with the emission dipole here. Further, the actual number of photons in the pixels of the image can be subjected to photon shot noise.

\begin{figure}[!htb]
    \centering
    \includegraphics[scale=0.5]{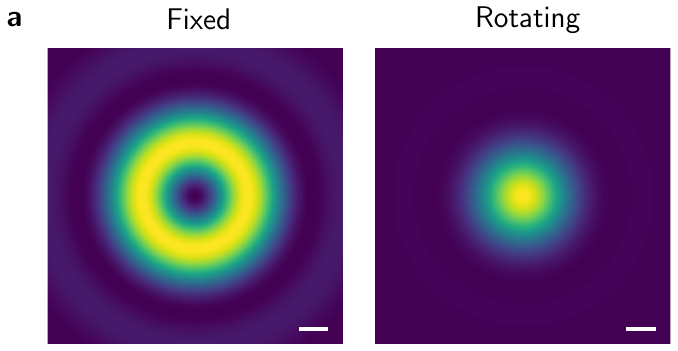}
    \includegraphics[scale=0.5]{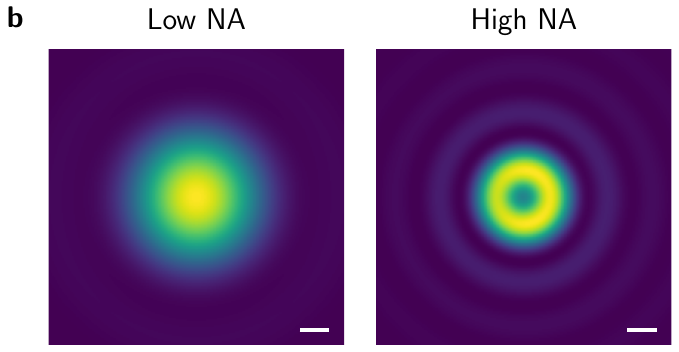}
    \vspace{1mm}
    \includegraphics[scale=0.5]{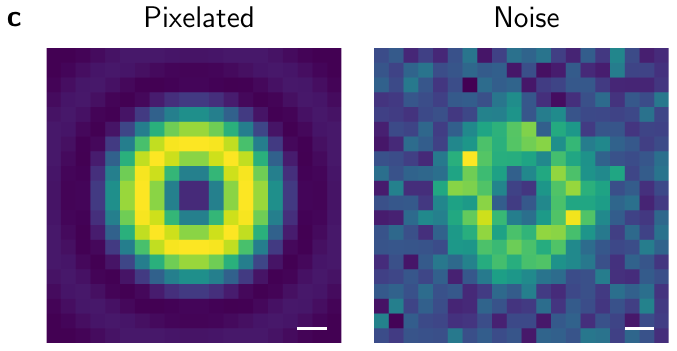}
    \includegraphics[scale=0.5]{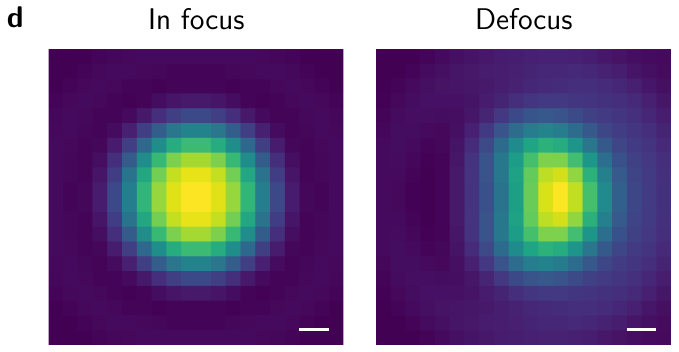}
    \caption{Fluorophore and microscope parameters. \panel{a} Fluorophore with a fixed dipole orientation of $(\theta,\phi)=(0,0)$ (left) and freely rotating fluorophore (right). \panel{b} PSFs shown for a fluorophore with dipole orientation of $(\theta,\phi)=(\frac{\pi}{8},0)$ for low and high objective numerical aperture (NA) with NA$=0.7$ (left) and NA$=1.2$ (right). \panel{c} Focal plane pixel size (in object space) of \SI{100}{\nano\meter} without noise (left) and with Poissonian background noise (right). Dipole orientation: $(\theta,\phi)=(0,0)$. \panel{d} PSF in focus (left) and PSF with an objective defocus of \SI{1}{\micro\meter} (right). Dipole orientation: $(\theta,\phi)=(\frac{\pi}{10},0)$. Scale bars: \SI{200}{\nm}}
    \label{fig_dipole_rotation}
\end{figure}

The configurable parameters for the microscope include settings of the tube lens (focal length) and objective (numerical aperture, magnification, focal length). The focus position of the objective is set to the transition between objective and intermediate layer. The focus position can be adjusted by changing the defocus value. In addition, an intermediate layer can be added, representing, e.g., a layer on the coverslip with a different refractive index than the immersion medium.

For the camera, the pixel size can be specified either by specifying the physical camera pixel size or the pixel size in object space. Further, background noise can be added to the image. The background noise is assumed to follow a Poisson distribution; the parameter for the background noise specifies its standard deviation.

\subsection{Multiple fluorophores}
\begin{figure}[!htb]
    \centering
    \includegraphics[scale=0.5]{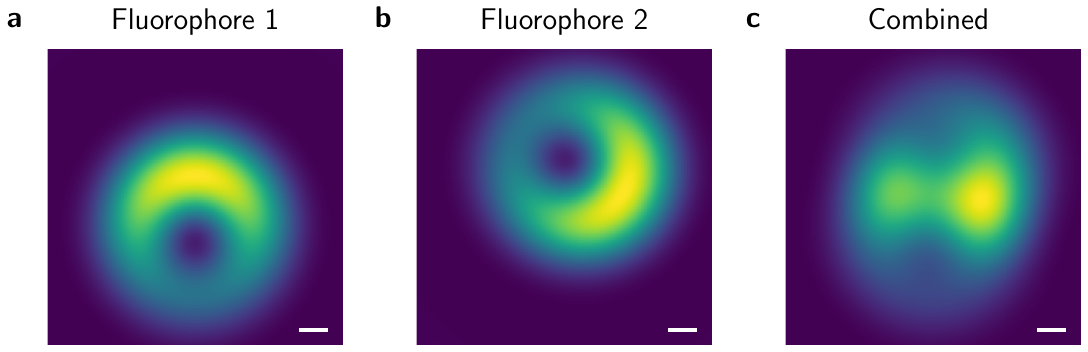}
    \caption{Multiple fluorophores. Panel \panel{a} and \panel{b} show individual emitters with different dipole orientations and positions. Fluorophore 1: $(\theta,\phi)=(\frac{\pi}{4},0)$, $(x,y,z)=(0,-200,0)$. Fluorophore 2: $(\theta,\phi)=(\frac{\pi}{4},\frac{2\pi}{3})$, $(x,y,z)=(100,200,0)$. Positions coordinates given in \SI{}{\nm}. \panel{c} Resulting additive PSF. Scale bars: \SI{200}{\nm}.}
    \label{fig:multiple_fluorophores}
\end{figure}

Ideally, the signals of active emitters in SMLM are well separated. However, this might not always be the case, in particular for high-density SMLM \cite{marsh2018} and stepwise photobleaching methods \cite{Gordon_2004}, where the PSFs of several fluorophores may overlap. In addition to single emitters, we allow to configure multiple fluorophores with different positions (lateral and axial) and dipole orientations in the same image.
The resulting PSF is the superposition of all individual PSFs and can show a distinctively different shape than the individual PSFs as shown in Fig.~\ref{fig:multiple_fluorophores}.

\subsection{Aberrations, transmission and phase retrieval}
\begin{figure}[!htb]
    \centering
    \includegraphics[scale=0.5]{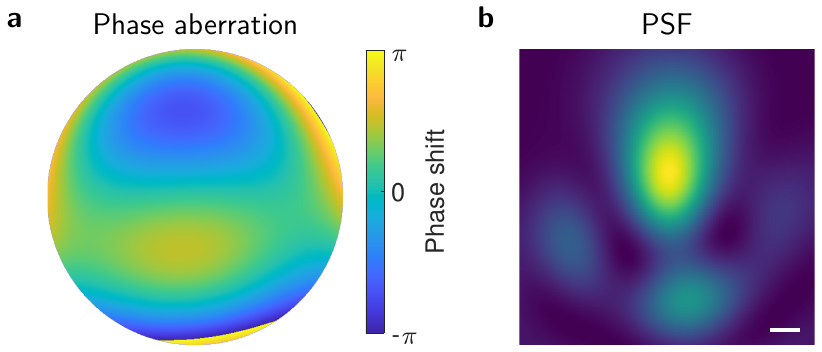}
    \caption{Aberrations. \panel{a} Aberration modeled by phase shift in the back focal plane. \panel{b} Resulting aberrated PSF. Scale bars: \SI{200}{\nm}.}
    \label{fig:aberrations}
\end{figure}

Up to now, we have assumed an ideal PSF only affected by noise. However, imperfections in the optical path or inhomogeneous refractive indices in the sample can lead to aberrated PSF shapes \cite{Ghosh_2015}. We allow to include such aberrations in the simulation by introducing a phase shift in the back focal plane that is expanded into a linear combination of Zernike polynomials (see Eq.~\ref{eq:zernike}). The most common aberrations, including spherical aberrations, astigmatism and coma can be directly selected and their coefficients adjusted. Alternatively, an arbitrary set of Zernike modes and corresponding coefficients can be specified. Fig.~\ref{fig:aberrations} shows an illustrative example how aberrations can affect the PSF shape.

In addition, the PSF shape can be affected by apodization, i.e.,\ non-homogeneous transmission of the emitted light through the objective. In particular, towards the outer rim of the objective, light transmission is reduced \cite{pawley2006handbook}. In order to model this attenuation, the app allows to load a custom transmission mask.

\begin{figure}[!htb]
    \centering
    \includegraphics[scale=0.5]{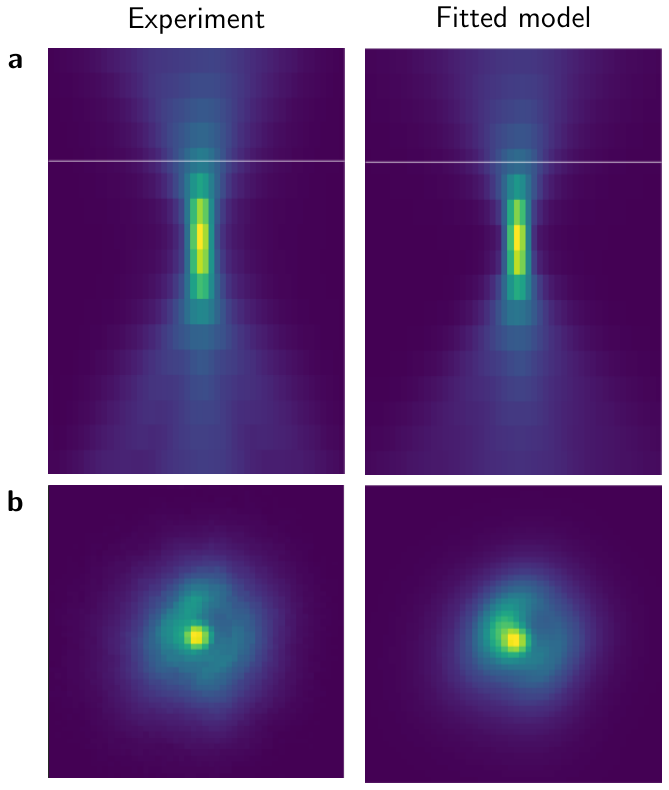}
    \caption{Phase retrieval. The left column shows the experimental input data, the left column the calculated model after fitting the aberrations. Projections of the PSF onto the xz-plane are shown in \panel{a}. Panel \panel{b} shows a 2D xy-view of the PSF at the defocus position indicated in panel \panel{a} by the white vertical line.}
    \label{fig:fitting}
\end{figure}

As an additional feature, we provide a subroutine that allows for retrieving the aberrations and transmission of a specific setup from experimental data. For this, a stack of images at various defocus positions from a calibration sample (a fluorescent bead) is required. For details on the recording of the calibration data see \ref{sec:phaseretrieval}. An illustrative example of experimental data and the fitted model is shown in Fig.~\ref{fig:fitting}.

\subsection{Phase masks}
\begin{figure}[!htb]
    \centering
    \includegraphics[width=\textwidth]{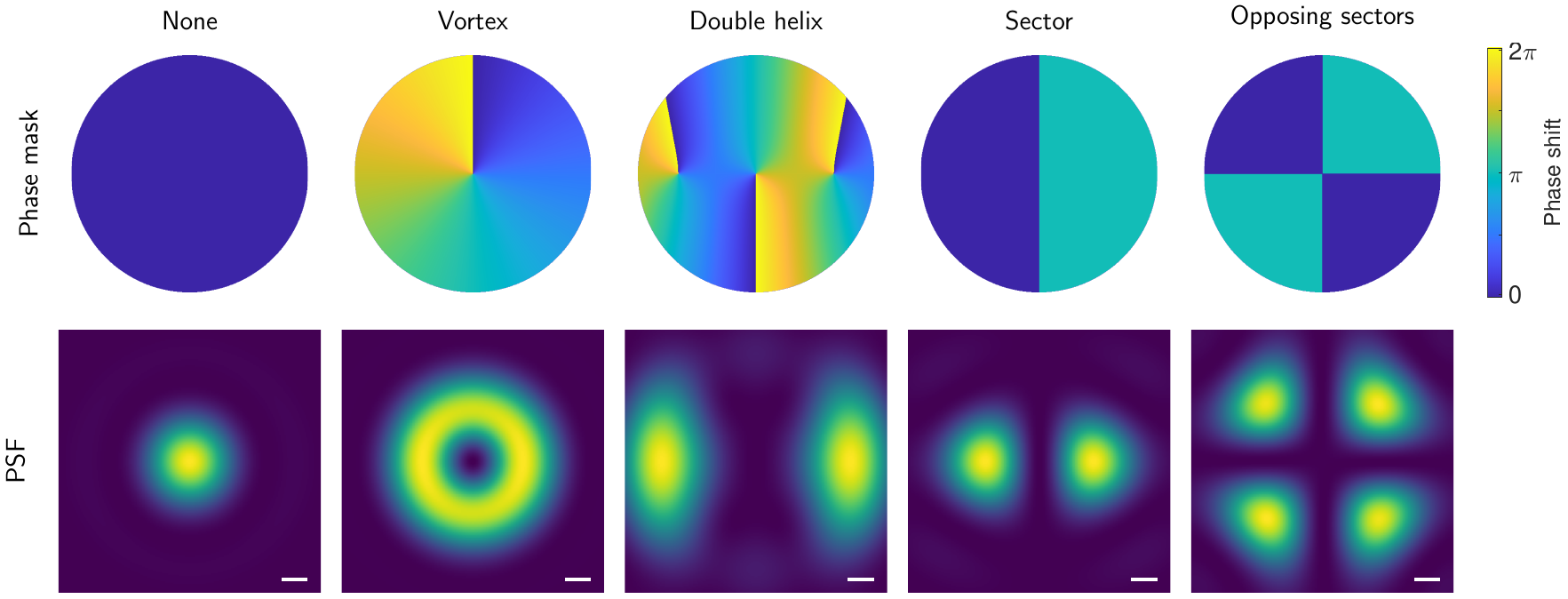}
    \caption{Phase masks. The top row shows the phase shift introduced by various phase masks and the bottom row the resulting PSF shape.}
    \label{fig:phasemasks}
\end{figure}

In contrast to undesired aberrations, phase manipulations can be used deliberately in PSF engineering approaches. Here, phase shifts are exploited for shaping the PSF in a way that allows to encode more information. For example, the double helix PSF has been used to allow for determination of the axial position \cite{pavani2009}, and a vortex phase mask has been shown to allow for retrieving information about the lateral and axial position, as well as the emitter's dipole orientation \cite{hulleman2021}. Any such phase manipulation can be introduced by adding an additional phase factor in the back focal plane (see Eq.~\ref{eq:aberrations}). Our app offers the feature to select from a variety of commonly used phase masks, including the vortex, double helix, and pyramid phase masks. In addition, a custom phase mask can be loaded. Further, the selected phase masks can be altered by cutting out an inner disk or rotation of the phase mask. A selection of phase masks and their influence on the PSF shape is given in Fig.~\ref{fig:phasemasks}.

\subsection{Visualization options}
Our app allows to visualize the calculated PSF for a set of given parameters in several ways. The default visualization option is a 2D lateral view of the PSF. In addition to the full PSF, the emission can be split into x- and y-polarization channels that can be viewed separately. Further, a full 3D model of the PSF can be calculated. For visualization, we show an xz-projection along with an isosurface plot. The value of the isosurface can be adjusted to show various isosurfaces of the 3D PSF.

\begin{figure}[!htb]
    \centering
    \includegraphics[scale=0.5]{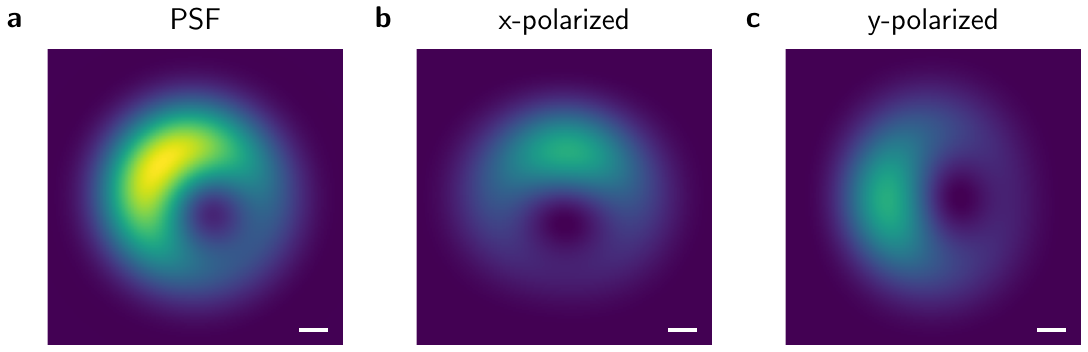} 
    \vspace{1mm}
    \includegraphics[scale=0.5]{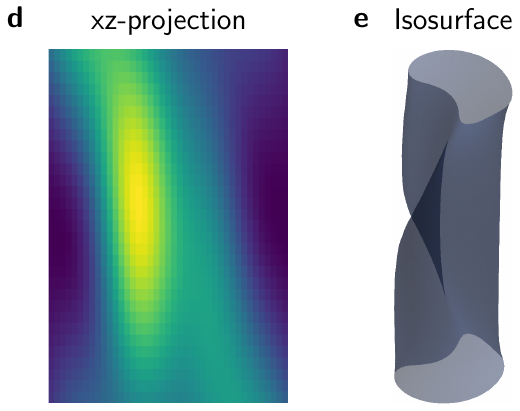}
    \caption{PSF visualization. The top row shows visualization options in 2D, including the 2D PSF \panel{a} and the split of the emission into polarized channels, \panel{b} and \panel{c}. The bottom row shows visualization options for the 3D PSF. \panel{d} xz-projection of the 3D PSF. \panel{e} Isosurface of 3D PSF. Scale bars: \SI{200}{\nm}.}
    \label{fig:psf_plots}
\end{figure}

\begin{figure}[!htb]
    \centering
    \includegraphics[scale=0.5]{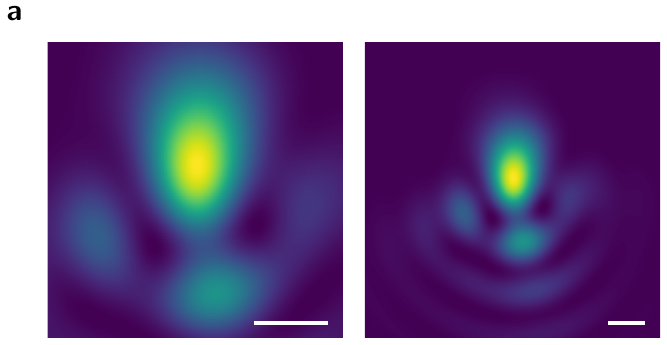}
    \includegraphics[scale=0.5]{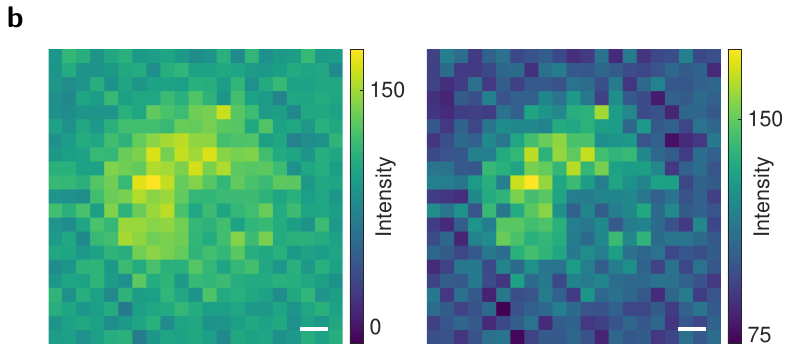}
    \vspace{1mm}
    \includegraphics[scale=0.5]{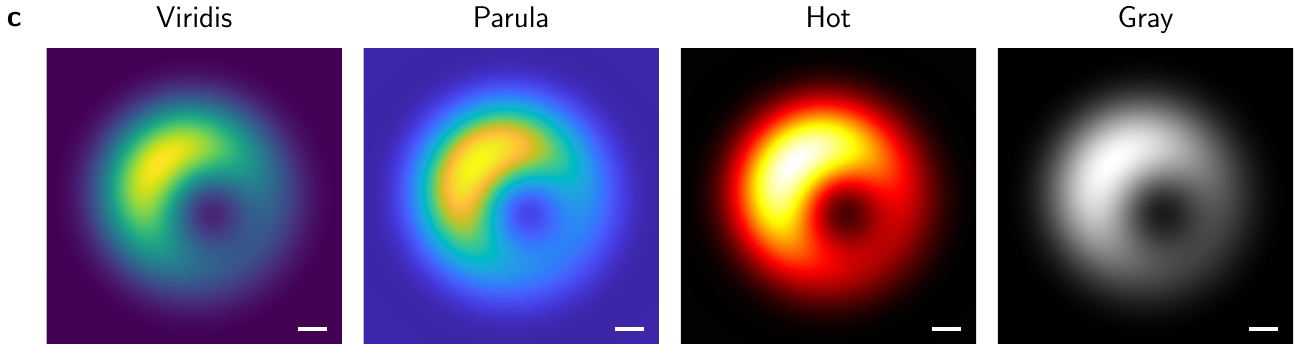}
    \caption{Visualization adjustments. \panel{a} Size of region of interest. \panel{b} Image contrast. Left: low contrast, right: high contrast. \panel{c} Colormap options. Scale bars: \SI{500}{\nm} (panel~a), \SI{200}{\nm} (panels~b-c).}
    \label{fig:plot_adjustments}
\end{figure}

The visualization of the plots can be adjusted in several ways as depicted in Fig.~\ref{fig:plot_adjustments}. First, the simulated region of interest can be specified by setting either the side length of the desired region of interest or the number of pixels per lateral axis (panel a). Second, in case of high background noise, adjusting the contrast of the image may help to better discern the PSF shape (panel b). Third, the colormap of the images can be set by selecting from a few options including the viridis, parula, hot and gray colormaps (panel c).

\subsection{Import and export options}
As noted in the previous subsections, our application allows to import custom data for aberrations, phase masks and transmission.
In the subwindow for fitting of the PSF model to experimental data, the fitting results (including aberrations and transmission) can be exported. The saved aberrations and transmission can then be imported into the main window via the aberration and transmission tabs. This allows to incorporate the fit results directly into the simulated model.
The calculated simulation data for 2D and 3D PSFs can be exported under the tab \textit{Options} for further processing or benchmarking of fitting algorithms.

\subsection{Cramér-Rao bound}
Finally, the theoretically achievable localization precision for lateral and axial position can be directly calculated in our app under the tab \textit{Options}. We calculate the CRB as the diagonal elements of the inverse Fisher information matrix (see Eq.~\ref{eq:crb}).
The CRB is affected by the shape of the PSF, the number of observed photons and the background noise. Note that in some cases the PSF does not provide enough information to estimate a particular parameter. For example, the axial position of a molecule that is in focus cannot be accurately estimated from an isotropic PSF in the absence of PSF engineering. In this case, numerical errors lead to unstable values of the CRB. Thus, we do not provide an exact value for the CRB  if the calculated number is greater than \SI{1000}{\nm}, i.e.,\ much greater than the diffraction limit of around \SI{200}{\nm}.

\section{Conclusion and outlook}
We have presented an easy-to-use MATLAB application that enables the simulation of point spread functions as they appear in SMLM applications. Reflecting the variability in experimental setups, the application offers a wide range of customizable parameters. This allows to tailor the simulation to specific microscope setups, ensuring that the simulated data closely aligns with the experimental reality.
The key features of the application include
\begin{itemize}
\item the simulation of fixed and rotating dipole emitters
\item the simulation of aberrations modelled by Zernike polynomials in the objective pupil
\item the addition of optical elements such as phase masks and polarizers in the emission path
\item retrieval of Zernike aberrations from an experimentally recorded PSF stack
\item the calculation of the Cramér Rao Bound to assess the theoretically achievable localization precision under specific conditions
\item flexible visualization options.
\end{itemize}

The results can be exported in various formats, allowing users to easily generate simulated data for an array of purposes. 
Most of the parameters can be adjusted via sliders or numerical input fields with a near real-time calculation and visualization of the PSF.
While we strived to offer computational efficiency for this first release of our app, a further speed up of calculations, in particular for 3D PSFs and small pixel size, can be achieved in future releases by leveraging GPU capabilities.
The features of this app and its interactivity allows users to observe the effect of specific parameters on the shape of the PSF, which, combined with the Cramér-Rao bound may assist, e.g., in designing novel PSF engineering approaches.


\section{Mathematical model}
We start by stating the PSF model for a fixed dipole emitter situated on the optical axis in an aberration-free optical system as illustrated in Fig.~\ref{fig:setup}. We assume that the emitter is embedded in a sample medium (e.g., water), followed by an intermediate layer (e.g., a transparent coating) and an objective layer (e.g., immersion oil). The interfaces are assumed to be planes orthogonal to the optical axis. 

The angles $(\theta,\phi)$ represent the inclination and azimuth angles characterizing the dipole orientation. We denote by $\lambda$ the emission wavelength in vacuum and by $\xb \in \mathbb{R}^2$ and $\xf \in \mathbb{R}^2$ coordinates in the back focal plane and image plane, respectively. 

The dipole emission pattern, interaction with the different layers of medium and subsequent passage through the infinity-corrected objective is comprehensively described in \cite{Axelrod_2012}. The starting point of our model is the electric field vector $E_{\text{BFP}}=\fieldBPF(\xb; \theta, \phi)$, expressed in Cartesian coordinates. This field in the back focal plane is defined by \cite[Eq. (18)]{Axelrod_2012}.

A tube lens with focal length $f$ is positioned between the objective and the camera to produce the image. The passage of the unaberrated field through this tube lens is modelled by the Fourier transform, 
\begin{equation}\label{eq:field}
     E(\xf) = \mathcal{F}(\fieldBPF)(\xf) := \frac{1}{i\lambda f}e^{\frac{2i\pi}{\lambda f}\|\xf\|}\int \fieldBPF(\xb) e^{-\frac{2 \pi i}{\lambda f}\xb \cdot \xf } d\xb \, .  
\end{equation}
Integration is performed over the circular pupil area. The intensity distribution in the focal plane of the tube lens is given by the absolute value of the electric field,
\begin{equation}\label{eq:PSF_intensity}
I(\xf) = |E(\xf)|^2.
\end{equation}
The coordinate system in the back focal plane can in principle be chosen arbitrarily. If one uses emission polarizers, the directions of $\xb$ are chosen to lie along the orthogonal directions of those polarizers. Then the field resulting from a linear polarizer oriented to transmit x-polarization (y-polarization) only is given by the first (second) component of $E(\xf)$ \cite{Axelrod_2012}.

\subsection{Discretization and noise}
The BFP field $\fieldBPF$ is calculated as implementation of \cite[Eqs. (10)-(18)]{Axelrod_2012} on a discrete grid. To simulate the pixelated grid of the camera, we consider the integrated intensity over the k-th pixel $\square_k$, 
\begin{equation}\label{PSF_intensity_pixel}
    I_{k} = \int_{\square_k} I(\xf) d\xf 
\end{equation}
We call the amount of supporting points used in each dimension for the calculation of \eqref{PSF_intensity_pixel} the \textit{oversampling factor}. 

The isotropic PSF resulting from a freely rotating emitter is modelled as the superposition of three fluorophores with pairwise orthogonal orientations. 

The intensity pattern resulting from multiple nearby emitters is calculated as the superposition of the respective individual intensities. 

Camera shot noise is modelled via the realization of a Poissonian random variable, with the calculated noise-free PSF as the mean. A constant background may be added before the application of the Poissonian random variable to model the fluorescence emission of a homogeneous background.

\subsection{Aberrations and phase shifts}
The infinity-corrected optical system provides a space between the objective and the tube lens, where additional optical components such as phase plates can be placed. Any wavefront deformation in this space, either caused by aberration or deliberate distortion, can be modeled by introducing additional phase factors.
\begin{equation} \label{eq:aberrations}
     E(\xf) = \mathcal{F}(e^{\frac{2i\pi}{\lambda}\varphi }\fieldBPF)(\xf) 
     \end{equation} 
We split the phase into two separate parts,
$\varphi = \varphi_{z} + \varphi_r$, where $\varphi_{z}$ is a Zernike term and $\varphi_{r}$ models the addition of further optical elements, e.g., phase masks in the back focal plane, see Section~\ref{section:features}. We expand $\varphi_{z}$ into a linear combination of orthonormal Zernike polynomials, i.e.,
    \begin{equation} \label{eq:zernike}
        \varphi_{z}(\xb) = \sum_{i}  w_i Z_i(\xb)\,, 
    \end{equation} 
where $Z_i$, denotes the $i$-th Zernike polynomial (using Noll's indices \cite{Noll_1976}) and $w_i$ is the corresponding Zernike coefficient. In particular, we use tip and tilt to model the PSF of an emitter that is laterally displaced from the optical axis. Setting $w_{1,2} = 1\lambda$ produces a lateral shift of $\frac{2}{NA}\lambda$ in horizontal and vertical direction, respectively. 

\subsection{Cramér-Rao Bound}
We calculate the Cramér Rao Bound (CRB), which is a tool from estimation theory that provides a benchmark for the achievable precision of an estimator \cite{Kay_1993}. The CRB is given by the diagonal elements of the inverse Fisher information matrix of the likelihood function. The likelihood function can be constructed from the forward model \eqref{PSF_intensity_pixel} and an appropriate noise model. As noise model we choose Poissonian noise, which is a good approximation for camera shot noise. Following \cite{Chao_2016}, we can then calculate the Fisher information as
    \begin{equation}\label{eq:crb}
        \mathcal{I}_{ij} 
         = \sum \limits_{k}   \frac{\partial I_{k}}{\partial \xi_i}     \frac{\partial I_{k}}{\partial \xi_j}   \frac{1}{ I_{k} }\,,
    \end{equation} 	
where the summation is over the pixels of the image. The parameter vector $\xi$ denotes the parameters that one wishes to estimate, which is typically just the lateral position. However, they could also include axial position or orientation. The Fisher information matrix is always a symmetric matrix with as many rows as the amount of parameters that are estimated.

\subsection{Phase retrieval}\label{sec:phaseretrieval}
The PSFs of real microscopes usually differ from the theoretical model. This is in one part explained by the design of the optical elements, in particular the objective lens. Although a modern microscope objective consists of many individual lenses, truly isoplanatic imaging performance cannot be obtained and significant amounts of astigmatism and coma appear at increasing distance from the optical axis. 
Other effects are known to introduce spherical aberrations, predominantly a refractive index mismatch between the sample buffer solution and objective immersion medium, but also too high or low environmental temperatures or age-related refractive index changes of the immersion oil.
Even when these aberrations are small, they can cause systematic errors in the molecule position estimates on the order of several tens to hundreds of microns. In order to avoid these errors, we include aberrations in the model via the Zernike term $\varphi_r$. 
The necessary Zernike coefficients are estimated by a phase retrieval algorithm~\cite{Zelger_2018} which operates on an experimental 3D image (\textit{z-stack}) acquired from a single small fluorescent bead. The algorithm finds Zernike coefficients that minimize the squared L2-Norm of a vectorial error metric $\epsilon$, which is defined as:
\begin{equation}
\epsilon_k = \frac{|E_k^\gamma|}{\sum_k |E_k^\gamma| } - \frac{|S_k^\gamma|}{\sum_k |S_k^\gamma| }.
\end{equation}
Here, $E$ and $S$ represent the experimentally recorded and simulated three-dimensional bead intensity images and $k$ the voxel index. The quantity $\gamma$ is a user-definable scalar value between 0 and 1 that influences the fit performance. Smaller values of $\gamma$ assign increased weight to voxels of lower intensity, e.g., those in out-of-focus planes. We discovered that a $\gamma$ value of 0.5 was effective in the tested cases, whereas a value of 1 led to inaccuracies as the algorithm stopped at non-ideal local minima. 

Appropriate z-stacks should be recorded at a maximum possible signal to noise ratio and contain about 10 widefield images covering an axial range from about \SI{-1}{} to \SI{1}{\um} around the axial bead position.  
Ideally, the bead is immersed in a mounting medium with a refractive index higher than the NA of the objective. This avoids the formation of a supercritical angle fluorescence (SAF) zone in the objective pupil, and also avoids complications caused by fluorophores located in different regions within the bead emitting different amounts of SAF.    
The diameter of the bead should not be larger than \SI{200}{\nm}.





\section*{Supplementary material}
The Supplementary material includes a table with all simulation parameters used for creating the PSF images for the manuscript figures. The app files and a detailed manual of the app are provided on GitHub under the following link: https://github.com/schneidermc/psf-simulation-app

\section*{Acknowledgements}
MCS was funded by Howard Hughes Medical Institute, Janelia Research Campus.
GJS and AJ were funded by the Austrian Science Fund (FWF): P-36022B.
GJS and MCS were funded by  the Austrian Science Fund (FWF): F6809-N34. 
FH was funded by the Austrian Science Fund (FWF): F6805-N34. 

\bibliographystyle{elsarticle-num} 
\bibliography{mybib}
\end{document}


\maketitle

\vspace{-50pt}
\section*{Simulation parameters}
The simulation parameters used for all figures in the main manuscript are shown in the table below.
The following parameters were the same in all simulations:
\begin{itemize}
    \setlength\itemsep{-0.2em}
    \item emission wavelength: $\lambda=\SI{680}{\nm}$
    \item reduced excitation: off
    \item objective focal length: $f_\text{obj}=\SI{3}{\mm}$
    \item tube lens focal length: $f=\SI{180}{\mm}$
    \item magnification: $60\!\!\mathrel{\scalebox{0.9}{$\times$}}$
    \item RI sample layer: $n_1=1.33$
    \item RI immersion medium: $n_3=1$
    \item intermediate layer: off 
    \item transmission mask: none 
    \item integrated intensity: $N_{\textrm{photon}} = 5000$.
\end{itemize}

\captionsetup{width=0.9\textwidth}
\renewcommand{\arraystretch}{1.2}
\begin{longtable}{|l|c|c|c|c|c|c|c|c|c|c|c|} 
\caption{Overview of all simulations parameters used for creating the figures in the main manuscript.
Notation: $(x,y,z)$ fluorophore position, $(\theta,\phi)$ fluorophore dipole orientation, NA objective numerical aperture, $d$ defocus (in \SI{}{\nm}), $p_{\textrm{im}}$ pixel size in object space (in \SI{}{\nm}), $n_{\textrm{px}}$ number of pixels per lateral axis, $b$ standard deviation of Poissonian background noise, S.n.\ shot noise, Aberr. aberrations\\
$^*$Zernike coefficients (Noll indexing): $w_5=0.05, w_6=0.1, w_7=0.15, w_{11}=0.03$.\\
$^{**}$ Values of defocus range from $\SI{-1500}{\nm}$ to $\SI{1500}{\nm}$ in steps of $\SI{100}{\nm}$ for the 3D PSF}\\
\hline
\textbf{Figure} & $(x,y,z)$ & $(\theta,\phi)$ & NA & $d$ & $p_{\textrm{im}}$ & $n_{\textrm{px}}$ & $b$ & S.n. & Aberr. & Phase mask\\ \hline
\endfirsthead

\hline
\textbf{Figure} & $(x,y,z)$ & $(\theta,\phi)$ & NA & $d$ & $p_{\textrm{im}}$ & $n_{\textrm{px}}$ & $b$ & S.n. & Aberr. & Phase mask\\ \hline
\endhead

Fig.~3a (left)  & $(0,0,0)$     & $(0,0)$               & 0.7 & 0   & 10  & 200 & 0   & no  & no & none  \\ \hline
Fig.~3a (right) & $(0,0,0)$     & rotating              & 0.7 & 0   & 10  & 200 & 0   & no  & no & none  \\ \hline
Fig.~3b (left)  & $(0,0,0)$     & $(\frac{\pi}{8},0)$   & 0.7 & 0   & 10  & 200 & 0   & no  & no & none  \\ \hline
Fig.~3b (right) & $(0,0,0)$     & $(\frac{\pi}{8},0)$   & 1.2 & 0   & 10  & 200 & 0   & no  & no & none  \\ \hline
Fig.~3c (left)  & $(0,0,0)$     & $(0,0)$               & 0.7 & 0   & 100 & 20  & 0   & no  & no & none  \\ \hline
Fig.~3c (right) & $(0,0,0)$     & $(0,0)$               & 0.7 & 0   & 100 & 20  & 100  & yes & no & none  \\ \hline
Fig.~3d (left)  & $(0,0,0)$     & $(\frac{\pi}{10},0)$  & 0.7 & 0   & 100 & 20  & 0   & no  & no & none  \\ \hline
Fig.~3d (right) & $(0,0,0)$     & $(\frac{\pi}{10},0)$  & 0.7 & 1000& 100 & 20  & 0   & no  & no & none  \\ \hline
Fig.~4a         & $(0,-200,0)$  & $(\frac{\pi}{4},0)$   & 0.7 & 0   & 10  & 200 & 0   & no  & no & vortex  \\ \hline
Fig.~4b         & $(100,200,0)$ & $(\frac{\pi}{4},\frac{2\pi}{3})$     & 0.7 & 0 & 10  & 200 & 0   & no  & no & vortex  \\ \hline
Fig.~5          & $(0,0,0)$ & $(\frac{\pi}{2},0)$               & 0.7 & 0       & 10  & 200 & 0   & no  & yes$^*$   & none  \\ \hline
Fig.~7          & $(0,0,0)$ & $(\frac{\pi}{2},0)$               & 0.7 & 0       & 10  & 200 & 0   & no  & no        & various  \\ \hline
Fig.~8a-c       & $(0,0,0)$ & $(\frac{\pi}{5},\frac{7\pi}{4})$  & 0.7 & 0       & 10  & 200 & 0   & no  & no & vortex  \\ \hline
Fig.~8d-e       & $(0,0,0)$ & $(\frac{\pi}{5},\frac{7\pi}{4})$  & 0.7 & $^{**}$  & 50  & 40  & 0   & no  & no & vortex  \\ \hline
Fig.~9a (left)  & $(0,0,0)$ & $(\frac{\pi}{2},0)$               & 0.7 & 0       & 10  & 200 & 0   & no   & yes$^*$   & none  \\ \hline
Fig.~9a (right) & $(0,0,0)$ & $(\frac{\pi}{2},0)$               & 0.7 & 0       & 10  & 400 & 0   & no   & yes$^*$   & none  \\ \hline
Fig.~9b (left)  & $(0,0,0)$ & $(\frac{\pi}{5},\frac{7\pi}{4})$  & 0.7 & 0       & 100 & 20  & 100 & yes   & no        & vortex  \\ \hline
Fig.~9b (right) & $(0,0,0)$ & $(\frac{\pi}{5},\frac{7\pi}{4})$  & 0.7 & 0       & 100 & 20  & 100 & yes   & no        & vortex  \\ \hline
Fig.~9c         & $(0,0,0)$ & $(\frac{\pi}{5},\frac{7\pi}{4})$  & 0.7 & 0       & 10  & 200 & 0   & no   & no        & vortex  \\ \hline
\end{longtable}

\clearpage
\vspace*{5pt}
\section*{Parameters for phase retrieval}
For phase retrieval for the test data set in Fig.~6. the following parameters were used:

\subsubsection*{Calibration sample:}
\begin{itemize}
    \setlength\itemsep{-0.2em}
    \item Bead diameter: $\SI{0.17}{\um}$
    \item z-increment: $\SI{0.2}{\um}$
    \item RI sample layer: $n_1=1.33$
    \item Emission wavelength: $\lambda=\SI{670}{\nm}$
\end{itemize}

\subsubsection*{Microscope parameters:}
\begin{itemize}
    \setlength\itemsep{-0.2em}
    \item Magnification: $67\!\!\mathrel{\scalebox{0.9}{$\times$}}$
    \item NA: $1.49$
    \item RI immersion: $n_3=1.518$
    \item Pixel size (physical): $\SI{6.5}{\um}$.
\end{itemize}

\subsubsection*{Fitting parameters:}
\begin{itemize}
    \setlength\itemsep{-0.2em}
    \item Number Zernikes: 21
    \item Iterations: 1000 (max.\ number)
\end{itemize}